\documentclass[preprint,12pt]{elsarticle}
\usepackage{epsfig}
\usepackage{amssymb}
\usepackage{amsmath}

\journal{Journal of Physics and Chemistry of Solids}

\begin{document}

\begin{frontmatter}

\title{THz optical response of Ba(Fe$_{1-x}$Ni$_x$)$_2$As$_2$ films analyzed within the three-band Eliashberg s$_\pm $-wave model}

\author[label1]{Yurii A. Aleshchenko \corref{cor1}}
\ead{aleshchenkoya@lebedev.ru}
\cortext[cor1]{Corresponding author}

\affiliation[label1]{organization={P.N. Lebedev Physical Institute, Russian Academy of Sciences},
            addressline={Leninskiy Prospekt 53},
            city={Moscow},
            postcode={119991},
            state={},
            country={Russia}}

\author[label1]{Andrey V. Muratov}

\author[label2]{Elena S. Zhukova}

\author[label2]{Lenar S. Kadyrov}

\author[label2]{Boris P. Gorshunov}

\affiliation[label2]{organization={Moscow Institute of Physics and Technology},
            addressline={Institutskiy per. 9},
            city={Dolgoprudny},
            postcode={141700},
            state={Moscow Region},
            country={Russia}}

\author[label3,label4]{Giovanni A. Ummarino}

\affiliation[label3]{organization={Istituto di Ingegneria e Fisica dei Materiali, Dipartimento di Scienza Applicata e Tecnologia, Politecnico di Torino},
            addressline={Corso Duca degli Abruzzi 24},
            city={Torino},
            postcode={10129},
            state={},
            country={Italy}}

\affiliation[label4]{organization={National Research Nuclear University MEPhI (Moscow Engineering Physics Institute)},
            addressline={Kashirskoe shosse 31},
            city={Moscow},
            postcode={115409},
            state={},
            country={Russia}}

\author[label1]{Ilya A. Shipulin}

\begin{abstract}
The uncertainty of the nature of the normal state and superconducting condensate of unconventional superconductors continues to stimulate considerable speculation about the mechanism of superconductivity in these materials. Of particular interest are the type of symmetry of the order parameter and the basic electronic characteristics of the superconducting and normal states. We report the derivation of temperature dependences of the superconducting condensate plasma frequency, superfluid density, and London penetration depth by measuring terahertz spectra of conductivity and dielectric permittivity of the Ba(Fe$_{1-x}$Ni$_x$)$_2$As$_2$ thin films with different Ni concentrations. A comprehensive analysis of the experimental data was performed in the framework of the simple three-band Eliashberg model under the assumption that the superconducting coupling mechanism is mediated by antiferromagnetic spin fluctuations. The results of independent experiments support the choice of model parameters. Based on calculations of the temperature dependences of superconducting gaps, we may conclude that the obtained results are compatible with the scenario, in which Ba(Fe$_{1-x}$Ni$_x$)$_2$As$_2$ is a multiband superconductor with s$_\pm $-wave pairing symmetry.
\end{abstract}


\begin{highlights}
\item Using terahertz spectroscopy the temperature dependences of the superconducting condensate plasma frequency, superfluid density and London penetration depth were obtained for the underdoped, optimally doped, and overdoped Ba(Fe$_{1-x}$Ni$_x$)$_2$As$_2$ thin films with Ni concentrations $x=0.035$, 0.05, and 0.08.
\item The experimental data were analyzed in the framework of the simple three-band Eliashberg s$_\pm $-wave model under the assumption that the superconducting coupling mechanism is mediated by antiferromagnetic spin fluctuations.
\item The temperature dependences of the superconducting gaps were calculated.
\item The choice of the model parameters was supported by the results of independent experiments.
\end{highlights}


\end{frontmatter}



\section{Introduction}

The discovery of high-temperature superconductivity in the iron-based superconductors (IBS) has received great attention over the past 15 years \cite{Kamihara,Chen1,Chen2,Paglione1,Stewart}. Among these materials, probably the most promising and intensively studied is the 122-type family (AFe$_2$As$_2$ with A = Ba, Sr, Ca, Eu), for which large and high-quality single crystals can be readily grown. The 122 parent compounds are antiferromagnetic metals for which a long-range antiferromagnetic order with a spin density wave (SDW) develops with temperature decrease down to $T_N\approx 138$ K (for, e.g., BaFe$_2$As$_2$), accompanied with a structural first-order phase transition from tetragonal to orthorhombic phase. Superconductivity in the IBS can occur through chemical doping, which leads to extra holes \cite{Rotter}, electrons \cite{Sefal}, or chemical pressure (for isovalent substitution) \cite{Kasahara}, or by applying external pressure \cite{Alireza}. The Fermi surface in the Ba122 family pnictides with electron doping consists of hole barrels near the $\Gamma $ point of the Brillouin zone and electron barrels near the M point \cite{Ideta}, where several superconducting (SC) gaps could develop below critical temperature ($T_c$). The widely accepted pairing state of these compounds is the fully gapped s$_\pm $ phase~\cite{Chubukov1,Mazin_spm}.

There is a general consensus that spin fluctuations play an important role in the formation of Cooper pairs in pnictides. These compounds are the most recent example of systems for which BCS theory fails~\cite{Dolgov1} and an Eliashberg approach is necessary to correctly describe their physics for the presence of a strong interband coupling.

Some of the most important superconductor properties, such as SC gap values, optical characteristics, temperature dependences of the London penetration depth ($\lambda _L(T)$) and superfluid density ($\rho _{SC}(T)$) can be obtained from the terahertz (THz) measurements.
However, as concerned the electron-doped Ba(Fe$_{1-x}$Ni$_x$)$_2$As$_2$, such studies have been previously performed only by us for the optimally doped  samples~\cite{Ummarino-A}. In addition, a large scatter of experimental order parameters ($\Delta $) for different compositions of Ba(Fe$_{1-x}$Ni$_x$)$_2$As$_2$ has been observed in the literature~\cite{Aleshchenko2021,Abdel,Zeng,Kuzmicheva1}.

In this paper, for the first time we compare the SC properties of the Ba(Fe$_{1-x}$Ni$_x$)$_2$As$_2$ underdoped ($x=0.035$), optimally doped ($x=0.05$), and overdoped ($x=0.08$) films probed by the THz spectroscopy and analyzed in the framework of the multiband Eliashberg model.

\section{Experiment}
The Ba(Fe$_{1-x}$Ni$_x$)$_2$As$_2$ films ($x=0.035$, 0.05, and 0.08) with the thicknesses of 100--150~nm were grown by pulsed laser deposition (PLD) method on double-polished (001) CaF$_2$ single-crystalline substrates using a KrF excimer laser with a 248 nm wavelength, pulse duration of 25~ns, and repetition rate of 7 Hz. The base pressure of the chamber was below $1\times 10^{-8}$~mbar and slightly increased to about $2\times 10^{-7}$~mbar during the deposition process. More details on the sample preparation can be found in~\cite{Richter1,Richter2,Shipulin1,Shipulin2}. The produced films have a mirror-like surface and similar epitaxial quality with growth orientation along the $c$ axis as the films described in~\cite{Shipulin2}. The composition of the films was explored by EDX measurements. The resistivity measurements of the films were performed by the standard four-probe technique in the Van der Paw scheme. The transition temperatures evaluated at 90$\%$ of the normal state resistance are 21.1~K, 21.6 K, and 10.3 K for $x=0.035$, 0.05, and 0.08, respectively. In the case when the critical temperature is determined as the temperature of maximum of the derivative of resistivity with respect to temperature, the $T_c$ values are the following: 20.27~K ($x=0.035$), 20.36~K ($x=0.05$), and 9.27~K ($x=0.08$). The temperature-dependent DC resistivity of the Ba(Fe$_{1-x}$Ni$_x$)$_2$As$_2$ films is shown in Fig.~1(a). Note narrow (about 2 K) transition to the SC state for all films and resistivities comparable to single crystals~\cite{Wang}, which attests the structural quality of the deposited films. In the case of the $x=0.035$ film, we can see a semiconductor-like $\exp (-\alpha /k_{B}T)$ behavior ($\alpha $ is a constant), since at low doping this material is more similar to the insulating parent compound. The similar behavior of resistivity occurs for high-temperature SC cuprates~\cite{Popov}. The magnetotransport measurements of the films were performed in a physical property measurement system PPMS-9 (Quantum Design) using external magnetic fields up to 9~T along the $c$-axis~\cite{Shipulin1}. The measured $R(T)$ curves recalculated to the temperature dependences of the resistivity in various magnetic fields are shown in Figs.~1(b)--1(d).

\begin{figure}[!htb]
\includegraphics[width=13.0cm]{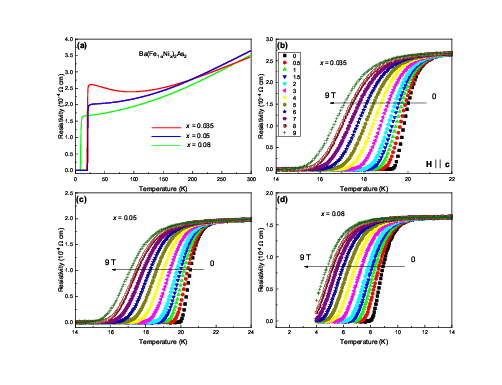}
\centering
\caption{(Color online) Temperature dependences of the resistivity for the Ba(Fe$_{1-x}$Ni$_x$)$_2$As$_2$ films without magnetic field (a) and in magnetic fields 0--9~T for $x=0.035$ (b), $x=0.05$ (c), and $x=0.08$ (d) (recalculated from the resistance curves~\cite{Shipulin1}).}
\end{figure}

THz spectroscopic measurements were carried out in the transmission mode with the Menlo pulsed time-domain THz spectrometer within the range of 10--50~cm$^{-1}$ (wavelengths 1 mm -- 200 $\mu $m) at different temperatures down to $T=4$ K with a home-made optical cryostat. Temperature-dependent THz spectra of the real and imaginary parts of the complex dielectric permittivity $\hat\varepsilon =\varepsilon _{1}(\omega)+i\varepsilon _{2}(\omega )$ and the complex optical conductivity $\hat\sigma (\omega )=\sigma _{1}(\omega )+i\sigma _{2}(\omega )$ were determined directly without using the Kramers-Kronig relations~\cite{Pracht}, by measuring the spectra of the complex transmission coefficient Tr$(\varepsilon _{1},\varepsilon _{2})$exp$[i\varphi ^{T}(\varepsilon _{1},\varepsilon _{2})]$ or Tr$(\sigma _{1},\sigma _{2})$exp$[i\varphi ^{T}(\sigma _{1},\sigma _{2})]$ of a two-layer system (a film on a substrate), see Eq. (6) in the Appendix. Using TeraCalc software, the numerical solution of the system of two essentially non-linear equations for the amplitude Tr$(\varepsilon _{1},\varepsilon _{2})$, or Tr$(\sigma _{1},\sigma _{2})$, and the phase $\varphi ^{T}(\varepsilon _{1},\varepsilon _{2})$, or $\varphi ^{T}(\sigma _{1},\sigma _{2})$, at each fixed frequency and temperature provided the spectra of the required optical parameters of the film. The dielectric properties of the CaF$_2$ substrate were measured beforehand.

\section{THz electrodynamics of Ba(Fe$_{1-x}$Ni$_x$)$_2$As$_2$ films}
To study the optical properties of the Ba(Fe$_{1-x}$Ni$_x$)$_2$As$_2$ films in detail, the response of charge carriers in the THz spectral range was analyzed, as in our previous paper~\cite{Ummarino-A}. The spectra of the real parts of dielectric permittivity $\varepsilon _1$ and conductivity $\sigma _1$ of the Ba(Fe$_{1-x}$Ni$_x$)$_2$As$_2$ films are shown in Figs.~2--4.

\begin{figure}[!htb]
\includegraphics[width=13.0cm]{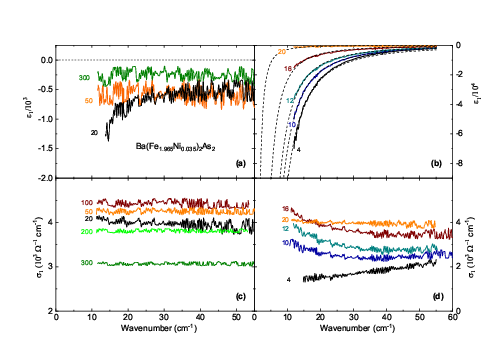}
\centering
\caption{(Color online) Spectra of the permittivity $\varepsilon _1$ and the conductivity $\sigma _1$ of the Ba(Fe$_{0.965}$Ni$_{0.035}$)$_2$As$_2$ films in the normal (a,c) and SC (b,d) states (numbers indicate temperature in Kelvins). Dashed lines in (b) show a fit of the spectra by the relation $\varepsilon _1\propto -(\omega _{p,s}/\omega )^2$.}
\end{figure}

The dependences for the normal state are presented in Figs.~2(a,c), 3(a,c), and 4(a,c), while those for the SC state are depicted in Figs.~2(b,d), 3(b,d), and 4(b,d). In the normal state, both $\varepsilon _1$ and $\sigma _1$ are nearly dispersionless and increase with cooling, thereby exhibiting metallic transport typical for the low-frequency limit (frequencies far below carrier scattering rate) of the Drude conductivity model~\cite{Sokolov}. For the films with $x=0.05$ and $x=0.08$ nickel contents, we observe at 20~K a slight decrease of conductivity with increasing frequency, indicating that the scattering rate approaches our operating frequency range from above~\cite{Sokolov}. In the SC state, THz conductivity of the films with Ni content $x=0.035$ and $x=0.05$ is strongly suppressed due to the opening of a SC gap. This suppression is accompanied by the emergence of a pronounced dispersion of permittivity shown in Figs.~2(b) and 3(b), which is a "Kramers-Kronig image" of the zero-frequency SC delta function in the conductivity spectra. This behavior is fitted by the expression $\varepsilon _{1}= -(\omega _{p,s}/\omega )^2$ (dotted lines), where $\omega _{p,s}$ is a SC plasma frequency. This fit allows us to determine $\omega _{p,s}$. Note that the conductivity of the $x=0.05$ film reaches nearly zero values around 10~cm$^{-1}$, while the conductivity of the $x=0.035$ film remains rather large, indicating a larger below-gap absorption. It is easily recognized that even at the lowest temperature (4~K) for all Ni contents the gapping is not complete. We have previously demonstrated~\cite{Aleshchenko2021} that, at least for the optimally doped Ba(Fe$_{0.95}$Ni$_{0.05})_2$As$_2$, only the narrow Drude component is gapped. According to~\cite{Lobo}, the presence of a low-frequency finite conductivity well into the SC state of iron pnictides could be also due to a gap anisotropy of the electron pocket~\cite{Chubukov2,Mishra,Carbotte1,Muschler}; impurity localized levels inside an isotropic SC gap~\cite{Shiba1,Shiba2,Rusinov1,Rusinov2,Schachinger}; or pair breaking due to interband impurity scattering in an s$_\pm $ symmetric gap~\cite{Vorontsov,Nicol}. We also observe a decrease below 100~K of the normal-state conductivity for $x=0.035$ film, which is related to the semiconductor-like $\rho (T)$ behavior discussed above.

\begin{figure}[htb]
\includegraphics[width=13.5cm]{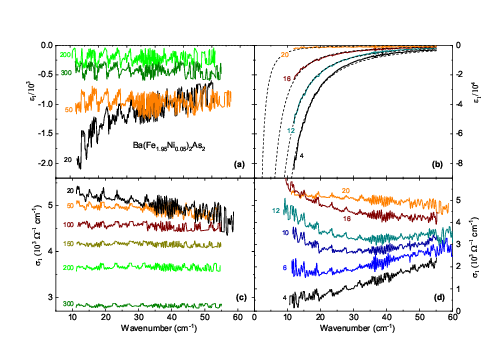}
\centering
\caption{(Color online) Spectra of the permittivity $\varepsilon _1$ and the conductivity $\sigma _1$ of the Ba(Fe$_{0.95}$Ni$_{0.05}$)$_2$As$_2$ films in the normal (a,c) and SC (b,d) states (numbers indicate temperature in Kelvins). Dashed lines in (b) show a fit of the spectra by the relation $\varepsilon _1\propto -(\omega _{p,s}/\omega )^2$.}
\end{figure}

\begin{figure}[h!tb]
\includegraphics[width=13.5cm]{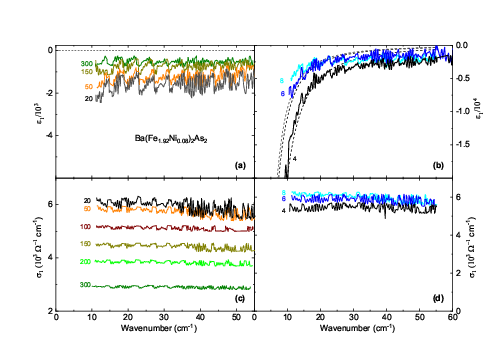}
\centering
\caption{(Color online) Spectra of the permittivity $\varepsilon _1$ and the conductivity $\sigma _1$ of the Ba(Fe$_{0.92}$Ni$_{0.08}$)$_2$As$_2$ films in the normal (a,c) and SC (b,d) states (numbers indicate temperature in Kelvins). Dashed lines in (b) show a fit of the spectra by the relation $\varepsilon _1\propto -(\omega _{p,s}/\omega )^2$.}
\end{figure}

In the case of the film with $x=0.08$, we can see at low temperatures a significant inductive response of the SC delta-function in the permittivity spectra (Fig.~4(b)). However, only a slight suppression of the conductivity is observed at the lowest attainable in our experiments temperature of 4~K (Fig.~4(d)), indicating that the suppression of conductivity due to the opening of a SC gap at $T=4$~K is below 10~cm$^{-1}$.

\begin{figure}[htb]
\includegraphics[width=12.0cm]{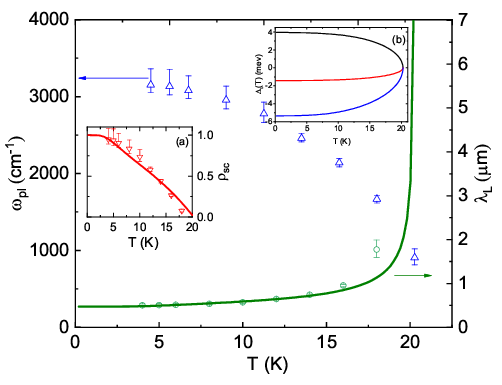}
\centering
\caption{(Color online) London penetration depth as a function of temperature for the Ba(Fe$_{0.965}$Ni$_{0.035}$)$_2$As$_2$ films (right scale): experimental data (open symbols) and theoretical calculations (solid green line). The temperature dependence of the plasma frequency of the SC condensate is also shown (left scale). The inset (a) shows the experimental temperature dependence of superfluid density $\rho _{sc}=[\lambda _L(0)/\lambda _L(T)]^2$ (open symbols) as well as the theoretical calculation (solid curve). The calculated SC gaps vs. temperature are presented in the inset (b). Here, the black curve represents the temperature dependence of the SC gap for the hole band, while the red and blue curves stand for the SC gaps of the two electron bands.}
\end{figure}

\begin{figure}[htb]
\includegraphics[width=12.0cm]{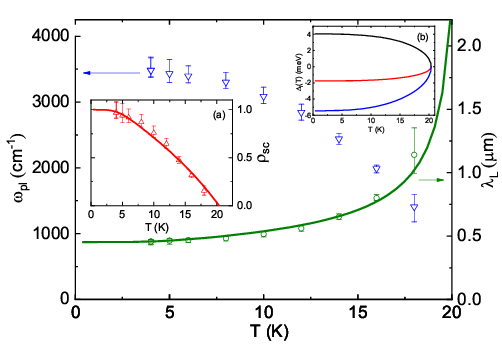}
\centering
\caption{(Color online) London penetration depth as a function of temperature for the Ba(Fe$_{0.95}$Ni$_{0.05}$)$_2$As$_2$ films (right scale): experimental data (open symbols) and theoretical calculations (solid green line). The temperature dependence of the plasma frequency of the SC condensate is also shown (left scale). The inset (a) shows the experimental temperature dependence of superfluid density $\rho _{sc}=[\lambda _L(0)/\lambda _L(T)]^2$ (open symbols) as well as the theoretical calculation (solid curve). The calculated SC gaps vs. temperature are presented in the inset (b). Here, the black curve represents the temperature dependence of the SC gap for the hole band, while the red and blue curves stand for the SC gaps of the two electron bands.}
\end{figure}

Figures 5, 6, and 7 show the calculated temperature dependences of the SC plasma frequency ($\omega _{p,s}$) and the London penetration depth ($\lambda _L$) (right scale) of the Ba(Fe$_{1-x}$Ni$_x$)$_2$As$_2$ films determined as $\lambda _{L}=c/\omega _{p,s}$~\cite{Basov} (where $c$ is the speed of light). The temperature dependence of the superfluid density $\rho_{sc}=[\lambda _{L}(0)/\lambda _{L}(T)]^2$ is shown in the insets (a) of Figs. 5 and 6. For the Ba(Fe$_{0.92}$Ni$_{0.08})_2$As$_2$ films, we failed to obtain a reliable experimental dependence of the $\rho_{sc}$ because of too low $T_c$.

\begin{figure}[htb]
\includegraphics[width=12.0cm]{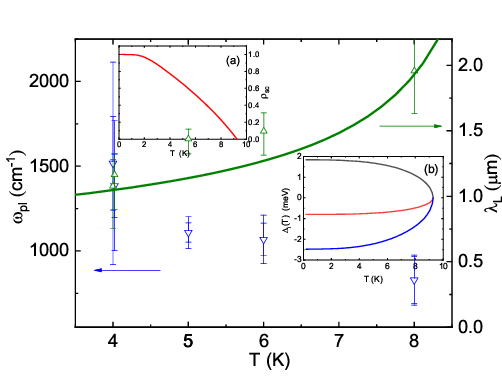}
\centering
\caption{(Color online) London penetration depth as a function of temperature for the Ba(Fe$_{0.92}$Ni$_{0.08}$)$_2$As$_2$ films (right scale): experimental data (open symbols) and theoretical calculations (solid green line). The temperature dependence of the plasma frequency of the SC condensate is also shown (left scale). The inset (a) shows the theoretical temperature dependence of superfluid density $\rho _{sc}=[\lambda _L(0)/\lambda _L(T)]^2$. The calculated SC gaps vs. temperature are presented in the inset (b). Here, the black curve represents the temperature dependence of the SC gap for the hole band, while the red and blue curves stand for the SC gaps of the two electron bands.}
\end{figure}

\section{Multiband Eliashberg theory}
We try to explain the experimental data in the framework of s$_\pm$ multiband Eliashberg theory \cite{Eliashberg,Chubukov,Carbotte, Ummamulti1,Ummamulti2}. Due to the fact that the electronic structure of Ba122 compounds doped with Ni or Co is extremely similar, it can be described in the framework of the three-band model~\cite{Ummaco}, which assumes the presence of one hole band and two electron bands. In this case, the hole band gap ($\Delta_{1}$) has opposite sign to the gaps of the two electron bands ($\Delta_{2}$ and $\Delta_{3}$). Despite the fact that this model is rather simple and has only two free parameters, it manages to reflect the basic physics of this class of compounds. To date, it is generally believed that in these materials the interband coupling between the hole and electron bands (s$_\pm$-wave model \cite{Aigor,Mazin_spm}) is provided mainly by antiferromagnetic spin fluctuations (\textit{sf}), while phonons can be responsible for the intraband coupling (\textit{ph})~\cite{Mazin_spm}. The calculation of basic superconducting parameters, such as the SC gaps as well as the critical temperature, can be done in the framework of the s$_\pm$-wave three-band Eliashberg equations, involving six coupled equations for the gaps $\Delta_{i}(i\omega_{n})$ and the renormalization functions $Z_{i}(i\omega_{n})$, where $i$ is a band index (that ranges between $1$ and $3$) and $\omega_{n}$ are the Matsubara frequencies. The imaginary-axis equations \cite{Umma1,Umma2,Umma3}, when the Migdal theorem works \cite{Migdal}, read:
\begin{eqnarray}
&&\omega_{n}Z_{i}(i\omega_{n})=\omega_{n}+ \pi T\sum_{m,j}\Lambda^{Z}_{ij}(i\omega_{n},i\omega_{m})N^{Z}_{j}(i\omega_{m})+\nonumber\\
&&+\sum_{j}\big[\Gamma _{ij}+\Gamma^{M}_{ij}\big]N^{Z}_{j}(i\omega_{n});
\label{eq:EE1}
\end{eqnarray}
\begin{eqnarray}
&&Z_{i}(i\omega_{n})\Delta_{i}(i\omega_{n})=\pi
T\sum_{m,j}\big[\Lambda^{\Delta}_{ij}(i\omega_{n},i\omega_{m})-\mu^{*}_{ij}(\omega_{c})\big]\times\nonumber\\
&&\times\Theta(\omega_{c}-|\omega_{m}|)N^{\Delta}_{j}(i\omega_{m})
+\sum_{j}[\Gamma _{ij}-\Gamma^{M}_{ij}]N^{\Delta}_{j}(i\omega_{n}),\phantom{aaaaaa}
\label{eq:EE2}
\end{eqnarray}
where $\Gamma _{ij}$ and $\Gamma^{M}_{ij}$ are the scattering rates from non-magnetic and magnetic impurities, respectively,
$$
\Lambda^{Z}_{ij}(i\omega_{n},i\omega_{m})=\Lambda^{ph}_{ij}(i\omega_{n},i\omega_{m})+\Lambda^{sf}_{ij}(i\omega_{n},i\omega_{m})
$$
and
$$
\Lambda^{\Delta}_{ij}(i\omega_{n},i\omega_{m})=\Lambda^{ph}_{ij}(i\omega_{n},i\omega_{m})-\Lambda^{sf}_{ij}(i\omega_{n},i\omega_{m}),
$$
where
\[\Lambda^{ph,sf}_{ij}(i\omega_{n},i\omega_{m})=2
\int_{0}^{+\infty}d\Omega\,\Omega\,
\alpha^{2}_{ij}F^{ph,sf}(\Omega)/[(\omega_{n}-\omega_{m})^{2}+\Omega^{2}]. \]
$\Theta$ is the Heaviside function and $\omega_{c}$ is a cutoff energy.  The quantities $\mu^{*}_{ij}(\omega _{c})$ are the elements of the $3\times 3$ Coulomb pseudopotential matrix. Finally, $N^{\Delta}_{j}(i\omega_{m})=\Delta_{j}(i\omega_{m})/
{\sqrt{\omega^{2}_{m}+\Delta^{2}_{j}(i\omega_{m})}}$ and
$N^{Z}_{j}(i\omega_{m})=\omega_{m}/{\sqrt{\omega^{2}_{m}+\Delta^{2}_{j}(i\omega_{m})}}$.
The electron-boson coupling constants are defined as
$$
\lambda^{ph,sf}_{ij}=2\int_{0}^{+\infty}d\Omega\frac{\alpha^{2}_{ij}F^{ph,sf}(\Omega)}{\Omega}.
$$
The solution of equations \ref{eq:EE1} and \ref{eq:EE2} requires a huge number of input parameters (18 functions and 27 constants), i.e.:
i) nine electron-phonon spectral functions $\alpha^{2}_{ij}F^{ph}(\Omega)$; ii) nine electron-antiferromagnetic spin fluctuation spectral functions, $\alpha^{2}_{ij}F^{sf}(\Omega)$; iii) nine elements of the Coulomb pseudopotential matrix $\mu_{ij}^{*}(\omega_{c})$;
iv) nine nonmagnetic $\Gamma _{ij}$ and nine paramagnetic $\Gamma^{M}_{ij}$ impurity-scattering rates.
However, some of these parameters can be extracted from experiments and some can be fixed by suitable approximations.
In particular, we refer to experimental data taken on high quality films, thus we can rather safely assume a negligible disorder: so we can put the interband scattering from nonmagnetic and magnetic impurities $\Gamma _{ij}$ and $\Gamma^{M}_{ij}$ equal to zero, while the nonmagnetic intraband scattering have no effect on critical temperature and gap values, so we can put here $\Gamma _{ij}=0$. For the term $\Gamma^{M}_{ii}$, we apply Occam's razor to explain the experimental data with the minimum number of starting hypotheses and assume that it is null. To escape this possibility would require, for example, measurements of the superconducting density of states obtained through tunneling. The presence of magnetic impurities unambiguously changes the shape of the density of states \cite{Golubovmag}.

At least as a starting point, let us make further assumptions that have been shown to be valid for iron pnictides \cite{Umma1,Umma2,Umma3}. Following ref.~\cite{Mazin_spm}, we can thus assume that: i) the total electron-phonon coupling constant is small (the upper limit of the phonon coupling in the usual iron-arsenide compounds is  $\approx0.35$ \cite{Boeri2}); ii) phonons mainly provide \emph{intra}band coupling so that $\lambda^{ph}_{ij}\approx0$; iii) spin fluctuations mainly provide \emph{interband coupling between hole and electron bands}, so that $\lambda^{sf}_{ii}\approx0$. Moreover, we put in first approximation the small phonon \emph{intra}band coupling $\lambda^{ph}_{ii}=0.1$, so as, following Mazin~I.I.~\cite{Aigor}, the Coulomb pseudopotential matrix: $\mu^{*}_{ii}(\omega _{c})=\mu^{*}_{ij}(\omega _{c})=0$
\cite{Umma1,Umma2,Umma3,Aigor}.
Within these approximations, the electron-boson coupling-constant matrix $\lambda_{ij}$ becomes:
\cite{Umma1,Umma2,Umma3}:
\begin{equation}
\vspace{2mm} %
\lambda_{ij}= \left (
\begin{array}{ccc}
  0.1                                                        &         \lambda^{sf}_{12}       &               \lambda^{sf}_{13}\\
  \lambda^{sf}_{21}=\lambda^{sf}_{12}\nu_{12}                &     0.1                         &                0            \\
  \lambda^{sf}_{31}=\lambda^{sf}_{13}\nu_{13}                &  0                              & 0.1                          \\
\end{array}
\right ), \label{eq:matrix}
\end{equation}
where $\nu_{ij}=N_{i}(0)/N_{j}(0)$ and $N_{i}(0)$ is the normal density of states at the Fermi level for the $i$-th band.
The coupling constants $\lambda_{ij}^{sf}$ are defined through the electron-antiferromagnetic spin fluctuation spectral functions (Eliashberg functions) $\alpha^2_{ij}F_{ij}^{sf}(\Omega)$. Following refs. \cite{Umma1,Umma2,Umma3} we choose these functions to have a Lorentzian shape, i.e.:
\begin{equation}
\alpha_{ij}^2F^{sf}_{ij}(\Omega)= C_{ij}\big\{L(\Omega+\Omega_{ij},Y_{ij})-
L(\Omega-\Omega_{ij},Y_{ij})\big\},
\end{equation}
where
\[
L(\Omega\pm\Omega_{ij},Y_{ij})=\frac{1}{(\Omega \pm\Omega_{ij})^2+Y_{ij}^2}
\]
and $C_{ij}$ are normalization constants, necessary to obtain the proper values of $\lambda_{ij}$, while $\Omega_{ij}$ and $Y_{ij}$ are the peak energies and the half-widths of the Lorentzian functions, respectively \cite{Umma3}.  In all these calculations, we set $\Omega_{ij}=\Omega_{0}$, thereby assuming that the characteristic energy of spin fluctuations is a single quantity for all the coupling channels and  $Y_{ij}= \Omega_{0}/2$, based on the results of inelastic neutron scattering measurements \cite{Inosov} (just for simplicity we choose the phonon spectral fuctions $\alpha_{ij}^2F^{ph}_{ij}(\Omega)$ with the same shape, same values of $\Omega_{ij}$ and $Y_{ij}$).

\begin{table*}
\centering
\begin{tabular}{|c|c|c|c|}
  \hline
  $x$   & $\lambda_{12}$ & $\lambda_{13}$ & $\lambda_{s,tot}$ \\
  \hline
  $x=0.035$ & 0.3100       & 1.4843       & 1.7967  \\
  $x=0.05$ & 0.4100       & 1.4845       & 1.8914  \\
  $x=0.08$ & 0.4100       & 1.4857       & 1.8926  \\

  \hline
\end{tabular}
\caption{Electron-boson coupling constants for different electron doping levels of Ba(Fe$_{1-x}$Ni$_x$)$_2$As$_2$ films}
\end{table*}

\begin{table*}
\centering
\begin{tabular}{|c|c|c|c|c|c|}
  \hline
   $x$ & $T_{c}$(K) & $\Delta_{1}$(meV) &$\Delta_{2}$(meV)&$\Delta_{3}$(meV) \\
  \hline
  0.035   & 20.27      & 4.24                & -1.78             &-5.90 \\
  0.05   & 20.36      & 4.27                & -1.78             &-5.93\\
  0.08   &  9.27      & 1.94                & -0.81             &-2.70\\

  \hline
\end{tabular}
\caption{Critical temperatures and theoretically calculated SC low temperature gaps of Ba(Fe$_{1-x}$Ni$_x$)$_2$As$_2$ films for different electron doping levels}
\end{table*}

The peak energy of the Eliashberg functions can be directly associated with the experimental $T_c$ by means of the empirical law $\Omega_{0}=2T_{c}/5$, which has been demonstrated to hold, at least approximately, for the IBS~\cite{Paglione1,Paglione}. With all these approximations, necessary to reduce the number of free parameters, this is the more simple model that can still grasp the essential physics of IBS. Within the framework of two-band models, experimental data can be reproduced, but electron-boson coupling constants have no direct physical interpretation ~\cite{Charnukha}. We use a cut-off energy $\omega_{c}=240$ meV and a maximum quasiparticle energy $\omega_{max}=250$~meV. The factors $\nu_{ij}$ that enter the definition of $\lambda_{ij}$ (eq. 3) are unknown, so we assume that they are equal to the Co-doped Ba122~\cite{Umma3}, so $\nu_{12}=1.12$ and $\nu_{13}=4.5$.
Now just two free parameters have to be determined $\lambda_{12}$ and $\lambda_{13}$ in the way to obtain the exact experimental critical temperature and the value of the small gap for $T<<T_{c}$. We take for the small gap of the Ba(Fe$_{1.965}$Ni$_{0.035}$)$_2$As$_2$ and Ba(Fe$_{1.95}$Ni$_{0.05}$)$_2$As$_2$ films the value close to that found in our previous infrared studies of the Ba(Fe$_{1.95}$Ni$_{0.05}$)$_2$As$_2$ films~\cite{Aleshchenko2021}. In the case of $x=0.08$, we vary $\lambda _{12}$ to obtain the experimental $T_{c}$. The obtained data are summarized in Tables~1 and 2. After this long discussion we arrived at a model with two free parameters. The best thing would be if they were calculated from first principles but this has not yet been done. The first conclusion, after determining these two parameters, is that the material is in a moderate strong coupling regime and these values are consistent with other similar materials.
We solve the imaginary-axis Eliashberg equations (eqs. \ref{eq:EE1} and \ref{eq:EE2}) to calculate the low-temperature values of the gaps, which are actually obtained by analytical continuation of the imaginary solutions to the real axis by using the technique of the Pad\'{e} approximants \cite{pade}.
The temperature dependence of $\Delta_{i}(i\omega_{n=0})$ is shown in the insets (b) of Figs. 5, 6 and 7, calculated by the solution of imaginary axis Eliashberg equations as well as the electron-boson spectral functions.

\section{Calculation of the penetration depth}
 The penetration depth, see Figs. 5, 6 and 7 (or the superfluid density, see relative insets (a)), can be computed starting from the gaps $\Delta_{i}(i\omega_{n})$ and the renormalization functions $Z_{i}(i\omega_{n})$ by~\cite{ghigolambda1,ghigolambda2}
\begin{eqnarray}
\lambda _{L}^{-2}(T)=(\frac{\omega_{p}}{c})^{2} \sum_{i=1}^{3}w_{i}\pi T \sum_{n=-\infty}^{+\infty}\frac{\Delta_{i}^{2}(\omega_{n})Z_{i}^{2}(\omega_{n})}{[\omega^{2}_{n}Z_{i}^{2}(\omega_{n})+
\Delta_{i}^{2}(\omega_{n})Z_{i}^{2}(\omega_{n})]^{3/2}}\,,\label{eq.lambda}
\end{eqnarray}
where $w_{i}=\left(\omega_{p,i}/\omega_{p}\right)^{2}$ are the weights of the single bands, $\omega_{p,i}$ is the plasma frequency of the $i$-th band, and $\omega_{p}$ is the total plasma frequency. Here, we can only act on the weights $w^\lambda_i$ in order to adapt the calculation to the experimental $\lambda_L(T)$. The multiplicative factor involving the plasma frequencies derives from the fact that the low-temperature value of the penetration depth $\lambda_L(0)$ should be related to the plasma frequency by $\omega_p=c/\lambda_L(0)$. In all cases, we found $w_{1}=0.5$, $w_{2}=0.3$, and $w_{3}=0.2$, while $\lambda _{L}(T=0)=0.58$~$\mu $m, $\lambda _{L}(T=0)=0.45$~$\mu $m, and $\lambda _{L}(T=0)=0.92$ $\mu $m for $x=0.035$, 0.05, and 0.08, respectively.

\section{Critical temperature versus magnetic field}

The multiband Eliashberg model developed above can also be used to explain the experimental results of upper critical field measurements \cite{Umma2,Sud,Bc2} as a function of temperature. For the sake of completeness, we give here the linearized gap equations in the presence of a magnetic field. In the following, $v_{Fj}$ is the Fermi velocity of band $j$ and $B_{c2}$ is the upper critical field:
\begin{eqnarray}
\omega_{n}Z_{i}(i\omega_{n})\hspace{-2mm}&\;=\;&\hspace{-2mm}\omega_{n}+\pi
T\sum_{m,j}[\Lambda_{ij}(i\omega_{n}-i\omega_{m})+\delta_{n,m}\frac{\Gamma _{ij}}{\pi T}]\mathrm{sign}(\omega_{m}),\nonumber\\
Z_{i}(i\omega_{n})\Delta_{i}(i\omega_{n})\hspace{-2mm}&\;=\;&\hspace{-2mm}\pi
T\sum_{m,j}\{[\Lambda_{ij}(i\omega_{n}-i\omega_{m})-\mu^{*}_{ij}(\omega_{c})]\times \nonumber\\
& &
\hspace{-2mm}\times\Theta(\omega_{c}-|\omega_{m}|)+\delta_{n,m}\frac{\Gamma _{ij}}{\pi T}\}\chi_{j}(i\omega_{m})Z_{j}(i\omega_{m})\Delta_{j}(i\omega_{m}),\nonumber
\end{eqnarray}

where

\begin{eqnarray}
\chi_{j}(i\omega_{m})&=&(2/\sqrt{\beta_{j}})\int^{+\infty}_{0}dq\exp(-q^{2})\times\nonumber\\
& & \hspace{-5mm}\times
\tan^{-1}[\frac{q\sqrt{\beta_{j}}}{|\omega_{m}Z_{j}(i\omega_{m})|+i\mu_{B}B_{c2}\mathrm{sign}(\omega_{m})}]\nonumber
\end{eqnarray}
with $\beta_{j}=\pi B_{c2} v_{Fj}^{2}/(2\Phi_{0})$.
In these equations, the three bare Fermi velocities $v_{Fj}$ are the input parameters. The number of adjustable parameters can be reduced \cite{Umma2} to one by assuming that, as in a free-electron gas, $v_{Fj}\propto N^{j} _{N}(0)$, so that $v_{F2}=\nu_{21} v_{F1}$ and $v_{F3}=\nu_{31} v_{F1}$.
We found $v_{F1}=8.9\times 10^{5}$~m/s if all impurities relative to electronic bands are in the band two ($\Gamma_{11}=912.3$~meV, $\Gamma_{22}=20.7$ meV and $\Gamma_{33}=0.0$ meV), while $v_{F1}=8.5\times 10^{5}$~m/s if all impurities relative to electronic bands are in the band three ($\Gamma_{11}=912.3$~meV, $\Gamma_{22}=0.0$ meV and $\Gamma_{33}=20.7$ meV). We have examined two extreme situations for the $x=0.05$ case. The real situation is probably among these two extreme cases. We fix the value of $B_{c2}$ and then  calculate $T_{c}$. Figure~8 demonstrates the results of the theoretical calculations, and it can be seen that in both cases the agreement with the experimental data is good for both $x=0.05$ (triangles) and $x=0.035$ (squares). The impurities related to electronic bands are all in the band 2 (red curve in Fig.~8) or in the band 3 (orange curve in Fig.~8). The values of $\Gamma _{ij}$ are determined by the fit of resistivity that will be shown in the forthcoming paper. The electron-boson spectral function normalized to unity in the case of $x=0.05$ is shown in the inset.

\begin{figure}[htb]
\includegraphics[width=12.0cm]{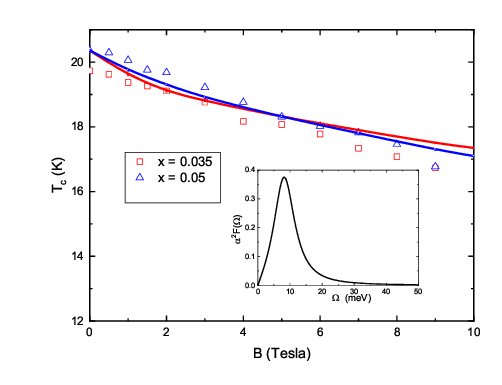}
\centering
\caption{(Color online) The critical temperature $T_{c}$ versus magnetic field $B$: experimental data (open symbols) and theoretical calculations (solid curves). The red and blue curves stand for the first and second cases, respectively (see the text), for $x=0.05$ doping. The electron boson spectral function normalized to one in the case of $x=0.05$ is shown in the inset.}
\end{figure}

\section{Discussion}

Within the framework of the three-band Eliashberg model, it was possible to describe in some detail the experimental temperature dependences of the superfluid density and penetration depth of the Ba(Fe$_{1-x}$Ni$_x$)$_2$As$_2$ films with different Ni concentrations. The calculated characteristic BCS ratios $\alpha =2\Delta /k_{B}T_c = 4.9$ and 6.8 of the larger gaps for the optimal composition ($x=0.05$) fall within the range reported in~\cite{Chi,Ding,Gong,Dressel,Kuzmicheva2,Kuzmicheva3}, the characteristic ratio of the smaller gap $\alpha =2.0$ correlates well with that found in~\cite{Kuzmicheva2,Kuzmicheva3}. We have also determined similar characteristic ratios in our recent optical studies of the optimally doped Ba(Fe$_{0.95}$Ni$_{0.05}$)$_2$As$_2$ films~\cite{Ummarino-A,Aleshchenko2021}.  Moreover, for the underdoped Ba(Fe$_{0.965}$Ni$_{0.035}$)$_2$As$_2$ films the characteristic ratios $\alpha =2.0$, 4.7, and 6.5 appear to be close to the characteristics for Ba(Fe$_{0.96}$Ni$_{0.04}$)$_2$As$_2$ single crystals~\cite{Sadakov}. It should be noted that in~\cite{Kuzmicheva1,Kuzmicheva2,Sadakov} three values of the order parameter were obtained and two larger ones were related to the possible anisotropic larger SC gap. Our theory predicts three SC gaps for all samples Ba(Fe$_{1-x}$Ni$_x$)$_2$As$_2$ ($x=0.035$, 0.05, and 0.08). However, in our previous work~\cite{Aleshchenko2021}, only two SC gaps were deduced from IR spectra of the Ba(Fe$_{0.95}$Ni$_{0.05}$)$_2$As$_2$films; it may be that one of the two gaps observed, usually the largest one, is a weighted average of the other two gaps predicted by the theory, while the smaller one is similar to that theoretically predicted.

\begin{figure}[htb]
\includegraphics[width=13.0cm]{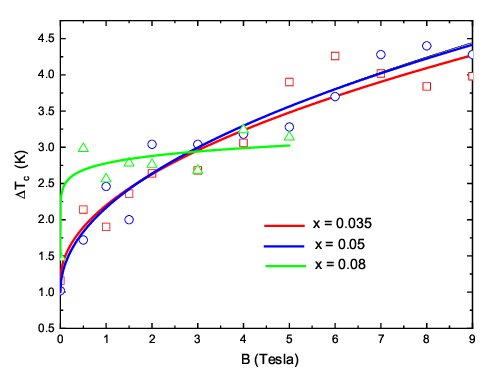}
\centering
\caption{(Color online) The width of the resistive transition $\Delta T_{c}$ as a function of the applied magnetic field for Ni contents $x=0.035$ (open red squares), $x=0.05$ (open blue circles), and $x=0.08$ (open green triangles). The fit with the function $\Delta T_{c}=a+bB^{\beta}$ is shown by the red solid line ($x=0.035$, $a=1.16000$, $b=1.03873$, $\beta =0.49922$), blue solid line ($x=0.05$, $a=1.02000$, $b=1.16648$, $\beta =0.48902$), and green solid line ($x=0.08$, $a=1.48000$, $b=1.29929$, $\beta =0.10741$).}
\end{figure}

Figure~9 shows the width of the resistive transition as a function of applied external magnetic field. The width of the resistive transition is defined~\cite{Tinkham1988} as $\Delta T_{c}=2(T_{c}-T_{c0})$, where $T_{c}$ is the critical temperature (the temperature where the derivative of the resistivity $d\rho(T)/dT$ has a maximum) and $T_{c0}$ is the maximum temperature, where $\rho(T)=0$.
It can be shown that for the high-temperature cuprates $\Delta T_{c} \sim B^{2/3}$~\cite{Tinkham1988}, assuming that the width of the zero magnetic field transition is negligible. If this does not happen, we try to reproduce the experimental data using a slightly more complex function of the form $\Delta T_{c}=a+bB^{\beta }$, where $a$, $b$, and $\beta $ are constants~\cite{Talantsev2021}. For samples with concentrations of $x=0.035$ and $x=0.05$, we found $\beta =0.499$ and $\beta =0.489$, respectively, which is close to $2/3$, while for $x=0.08$ we found $\beta =0.107$, which diverges significantly from theoretical predictions. It is interesting to note that optimally doped Ba(Fe$_{1-x}$Ni$_x$)$_2$As$_2$ films have a quite similar power dependence with the high-temperature cuprates.

\section{Conclusions}

In this work, we have performed detailed optical conductivity measurements for the epitaxial Ba(Fe$_{1-x}$Ni$_x$)$_2$As$_2$ thin films  ($x=0.035$, 0.05, and 0.08) on CaF$_2$ substrates in the normal and SC states using THz spectroscopy. Although a clear sign of the SC gap appears in the spectra, the conductivity does not vanish in the SC state for all Ni concentrations. A comprehensive analysis of this feature, as well as a description of the behavior of the temperature dependences of the superfluid density and the London penetration depth, was carried out in the framework of the three-band Eliashberg model. Moreover, the temperature dependences of the SC gaps were
calculated on the basis of this model, which allowed us to study the features of the SC state of this compound in more details. Various  independent experimental data (the temperature dependences of the upper critical magnetic field) support the choice of input parameters in the Eliashberg equations. In addition, our results are compatible with the scenario in which Ba(Fe$_{1-x}$Ni$_x$)$_2$As$_2$ over a wide range of Ni concentrations remains a multiband superconductor with s$_\pm $-wave pairing symmetry mediated mainly by antiferromagnetic spin fluctuations.

\section*{CRediT author statement}

\textbf{Yurii A. Aleshchenko}: Conceptualization, Writing - Original Draft. \textbf{Andrey V. Muratov}: Data Curation, Formal analysis, Visualization. \textbf{Elena S. Zhukova}: Investigation. \textbf{Lenar S. Kadyrov}: Investigation. \textbf{Boris P.Gorshunov}: Writing - Original Draft, Writing - Review \& Editing, Funding acquisition. \textbf{Giovanni A. Ummarino}: Methodology, Investigation, Writing - Original Draft, Visualization. \textbf{Ilya A. Shipulin}: Resources, Investigation, Writing - Review \& Editing.

\section*{Declaration of competing interest}

The authors declare that they have no known competing financial interests or personal relationships that could have appeared to influence
the work reported in this paper.

\section*{Data availability}

Data will be made available on request.

\section*{Acknowledgements}

We thank Ruben H\"uhne for fruitful discussions and providing samples. G.A. Ummarino acknowledges partial support from the MEPhI. THz experiments were supported by the Ministry of Science and Higher Education of the Russian Federation (No. FSMG-2021-0005),  THz data processing and analysis was carried out with the support of the Ministry of Science and Higher Education of the Russian Federation (Grant No. 075-15-2024-632).

\section*{Appendix}

The complex transmission coefficient for the two-layer system can be evaluated using the following equation~\cite{Dressel G}:
\begin{equation}
T_{1234}^*=\frac{T_{12}T_{23}T_{34}e^{i(\delta_2 +\delta _3)}}{1+R_{23}R_{34}e^{2i\delta _3}+R_{12}R_{23}e^{2i\delta _2}+R_{12}R_{34}e^{2i(\delta _2 +\delta _3)}} ,
\end{equation}
where $T_{pq}$ and $R_{pq}$ are Fresnel transmission and reflection coefficients, respectively, at the boundary between layers with indexes $p$ and $q$; $\delta _p=2\pi d_p(n_p+ik_p)/\lambda $ for $p=2$, 3, with $d_p$ being the thickness of the layer $p$; $n_p$ and $k_p$ are refraction index and extinction coefficient, respectively, of the layer $p$; $\lambda $ is the radiation wavelength. Namely, these coefficients can be written in exponential form, $T_{pq}=t_{pq}\exp (i\varphi _{pq}^T)$ and $R_{pq}=r_{pq}\exp (i\varphi _{pq}^R)$, and expanded through refraction index and extinction coefficient of the adjacent layers~\cite {Dressel G,Born}:
$$
t_{pq}^2=\frac {4(n_p^2+k_p^2)}{(k_p+k_q)^2+(n_p+n_q)^2} ;\quad r_{pq}^2=\frac {(n_p-n_q)^2+(k_p-k_q)^2}{(k_p+k_q)^2+(n_p+n_q)^2} ,
$$
$$
\varphi _{pq}^T={\rm arctg}\left(\frac{k_pn_q-k_qn_p}{n_p^2+k_p^2+n_pn_q+k_pk_q}\right) ;\quad\varphi _{pq}^R={\rm arctg}\left[\frac {2(k_pn_q-k_qn_p)}{n_p^2+k_p^2-n_q^2-k_q^2}\right] .
$$

\end{document}